# Spectroscopy of two-dimensional interacting lattice electrons using symmetry-aware neural backflow transformations

Imelda Romero*, Jannes Nys, and Giuseppe Carleo
*Institute of Physics, École Polytechnique Fédérale de Lausanne (EPFL), CH-1015 Lausanne, Switzerland and Center for Quantum Science and Engineering, École Polytechnique Fédérale de Lausanne (EPFL), CH-1015 Lausanne, Switzerland* *

Neural networks have shown to be a powerful tool to represent the ground state of quantum many-body systems, including fermionic systems. However, efficiently integrating lattice symmetries into neural representations remains a significant challenge. In this work, we introduce a framework for embedding lattice symmetries in fermionic wavefunctions and demonstrate its ability to target both ground states and low-lying excitations. Using group-equivariant neural backflow transformations, we study the $t$-$V$ model on a square lattice away from half-filling. Our symmetry-aware backflow significantly improves ground-state energies and yields accurate low-energy excitations for lattices up to $10 \times 10$. We also compute accurate two-point density-correlation functions and the structure factor to identify phase transitions and critical points. These findings introduce a symmetry-aware framework important for studying quantum materials and phase transitions.

## I. INTRODUCTION

Strong correlations lead to rich physical phenomena in quantum many-body systems, such as metal-insulator transitions, spin-charge separation, and the paradigmatic fractional quantum Hall effect [1–4]. The strong interactions among particles in these systems make their description complex. Various numerical methods have been developed to tackle the strongly correlated regime, including variational approaches such as variational Monte Carlo (VMC) [5] and tensor network methods [6]. Machine learning has recently found its application in quantum many-body physics to introduce flexible and powerful parameterizations of quantum states. This is guided by the capacity of neural networks to act as universal and efficient high-dimensional function approximators [7]. They have shown great potential, often resulting in state-of-the-art ground state approximations, especially in 2D [8–13], and have also found their application in dynamics [7, 14–20].

Neural network quantum states (NQS) [7] have also been used to simulate fermionic systems in the first and second quantization formalisms [21–27]. In the latter, the fermionic anticommutation relations make variational approaches challenging. This is particularly clear when mapping fermionic operators onto spin operators, e.g. using Jordan-Wigner in >1D, where these mappings introduce a highly non local spin Hamiltonian [28]. On the other hand, in the first quantization formalism one must exactly fulfill the particle-permutation antisymmetry of the wave function. A conventional variational wavefunction typically involves a (mean-field) Slater determinant to account for antisymmetry, combined with a two-body Jastrow factor [29] to capture particle correlations. One way to further improve this ansatz is by introducing correction terms known as backflow transformations (BF). This modification involves making the orbitals within the Slater determinant depend on the positions of all fermions. Feynman and Cohen originally introduced the idea to analyze the excitation spectrum of liquid Helium-4 [30], and it was successfully extended to electronic degrees of freedom [31–35]. Backflow transformations can alter the nodal surface, thereby reducing approximation errors [32, 36–38]. Recently, the backflow transformation has been introduced as a neural network in the context of NQS applied to discrete [39] and continuous [26, 27] fermionic systems. For spin degrees of freedom, it has been demonstrated that embedding symmetries into NQS can greatly improve ground state accuracy [13, 21, 40–46]. Furthermore, restoring the symmetries of the system enables us to target low-lying excited states that can be classified by the different symmetry sectors [21, 40]. One general way to target the low-energy states of the symmetry sectors is by applying quantum-number projectors to the wave function [21].

In this work, we introduce a method for embedding lattice symmetries of 2D fermionic lattice Hamiltonians into neural backflow transformations and demonstrate its efficacy using Slater-Backflow-Jastrow wavefunction ansatzes. Our approach incorporates symmetry-aware neural backflow transformations, fulfilling equivariance conditions for translational and particle-permutation symmetries through convolutional neural networks (CNN). By employing quantum number projection, we symmetrize the wavefunction and accurately target low-lying excited states by varying quantum numbers such as total momentum.

We benchmark our ansatz on the $t$-$V$ model on a square lattice and find that it significantly increases the ground-state accuracy compared to other state-of-the-art approaches. Additionally, it enables precise determination of low-lying excited states over a wide range of interaction strengths. The results demonstrate the robustness of this approach in capturing phase transitions and iden-tifying the critical interaction strength. Furthermore, a V-score analysis highlights the improved variational accuracy of our method across diverse interaction strengths, system sizes, and excitations, showcasing the effectiveness of symmetry-aware neural backflow transformations for fermionic systems.

* imelda.romero@epfl.ch

## II. RESULTS

### A. Hamiltonian and Observables

The Hamiltonian of the $t$-$V$ model reads

$$\hat{H} = -t \sum_{(i,j)\in\mathcal{E}} \hat{c}_i^\dagger \hat{c}_j + \hat{c}_j^\dagger \hat{c}_i + V \sum_{(i,j)\in\mathcal{E}} \hat{n}_i \hat{n}_j. \tag{1}$$

The first term describes electron hopping between neighboring sites with hopping parameter $t$. The second term corresponds to the nearest neighbor Coulomb repulsion with interaction strength $V \geq 0$. We will set $t = 1$ from hereon. We further decompose the Hilbert space $\mathcal{H}$ into fixed particle-number subspaces $\mathcal{H}_{N_f}$ [47]. The $t$-$V$ model was originally introduced to study the thermodynamic and transport properties of superconductors [48, 49]. Additionally, it provides a conceptual framework for explaining phenomena such as phase separation or stripe order in cuprates and organic conductors [50–55]. In practice, the $t$-$V$ model can be realized, for example, in experiments with strongly polarized $^3$He atoms [56, 57], or using cold atoms in optical lattices with Rydberg dressing [58]. Despite its apparent simplicity, the $t$-$V$ model cannot be solved analytically in two or higher dimensions. It also reveals highly nontrivial phase transitions that have been studied in previous works with various computational techniques, including variational Monte Carlo [22, 50, 59–66].

In the strong-coupling limit where $V/t \to \infty$ we encounter a charge-ordering (CO) phase where the system behaves classically. At large $V/t$, it is energetically unfavorable for two fermions to occupy neighboring lattice sites. The correlations become short-ranged, suggesting localization and insulating behavior, giving rise to a charge-ordered insulating phase. At half-filling, where the particle density is $\bar{n} = 0.5$, the charge order forms a checkerboard pattern. Conversely, at weak interaction strengths $V/t$, the fermions can easily hop between neighboring lattice sites, and the system behaves like a non-interacting free Fermi gas. In this weak coupling limit the system enters the metallic phase and becomes exactly solvable at $V/t \to 0$ [22, 59, 60, 67–72].

We introduce the normalized density-density correlation function as [22]

$$C(R) = \frac{1}{|\mathcal{V}||S_R|} \sum_{i\in\mathcal{V}} \sum_{j\in S_R(i)} \langle (\hat{n}_i - \bar{n})(\hat{n}_j - \bar{n}) \rangle \tag{2}$$

where $S_R(i) = \{j \in \mathcal{V} : d(i,j) = R\}$ is the set of vertices with a fixed distance $R$ from the vertex $i$. Another important observable to detect the CO phase is the structure factor [59, 60, 62, 63, 73]:

$$S(\mathbf{K}) = \frac{1}{N} \sum_{j,k\in\mathcal{V}} e^{i\mathbf{K}\cdot(\mathbf{r}_j - \mathbf{r}_k)} \langle (\hat{n}_j - \bar{n})(\hat{n}_k - \bar{n}) \rangle . \tag{3}$$

In the CO phase, well-defined peaks at $\mathbf{K} = (\pi, \pi)$ indicate a checkerboard charge pattern, and in the thermodynamic limit $S((\pi,\pi))/N_s$ converges to a finite value, reflecting long-range order.

We will study the $t$-$V$ model on a two-dimensional square lattice of side $L$ and with periodic boundary conditions, for various system sizes $|\mathcal{V}| = L^2$ and different interaction strengths $V/t$, with densities close to half-filling and at closed momentum shells.

### B. Ground States

We benchmark our symmetrized Neural Slater-Backflow-Jastrow ($\psi_{BF}^K$) ansatz with respect to a mean-field Slater determinant (which is equivalent to Hartree-Fock (HF)), a symmetrized Slater-Jastrow ($\psi^K$) without backflow, and a non-symmetrized Slater-Jastrow. Furthermore, we compare our results with ground state energies obtained using another state-of-the-art neural quantum state method, specifically the "Slater-Jastrow with an additional sign correction neural network" ansatz from Ref. [22]. In their method, they do not use the symmetry-averaging process. Instead, they employ a Slater-Jastrow inspired ansatz with deep residual networks and convolutional residual blocks to approximately determine the ground state of spinless fermions on a square lattice with nearest-neighbor interaction. The ansatz is a modified Hartree-Fock wavefunction, enhanced by a Jastrow factor and a sign-correcting neural network, both of which are constructed to be invariant under certain lattice symmetries.

It is important to note that the simulations in this section for the different system sizes are performed in an open shell domain, where fermions occupy the lowest-energy available states. In the context of spinless fermions, a “closed shell” configuration occurs when all momentum states within the Fermi surface are fully occupied, typically resulting in a non-degenerate ground state with total momentum $\mathbf{K} = (0, 0)$. In contrast, an “open shell” configuration arises when some momentum states remain unfilled, leading to a partially filled Fermi surface and potentially different total momenta for the ground state. Thus, in our open shell setup, the ground state does not necessarily correspond to $\mathbf{K} = (0, 0)$.

In Fig. 1 (a), we present results for a small system size $L = 4$ at a density of $\bar{n} = 0.44$ and ground state momentum $\mathbf{K} = (\pi, 0)$, allowing comparison to results obtained by exact diagonalization (ED). Mean-field Slater relative errors range from $10^{-3}$ to $10^{-1}$, with accuracies decreasing at large interaction strengths $V/t$. Our symmetrized backflow ansatz consistently yields ground-state errors below $10^{-3}$ , also at higher values of $V/t$, and achieves lower errors compared to other state-of-the-art methods. We also observe that the symmetry-aware backflow transformation yields the most accurate ground-state energies over the whole interaction regime. For a larger system size $L = 10$, at a density of $\bar{n} = 0.44$ and ground state momentum $\mathbf{K} = (\frac{3\pi}{5}, \frac{2\pi}{5})$, shown in Fig. 1 (b), this trend is confirmed. The backflow corrections significantly lower the estimated ground-state energy across all coupling strengths. We compare the converged VMC energies with HF energies by computing the difference $E - E_{\mathrm{HF}}$. Our results show that the backflow ansatz consistently provides lower energies across the full range of interaction strengths.

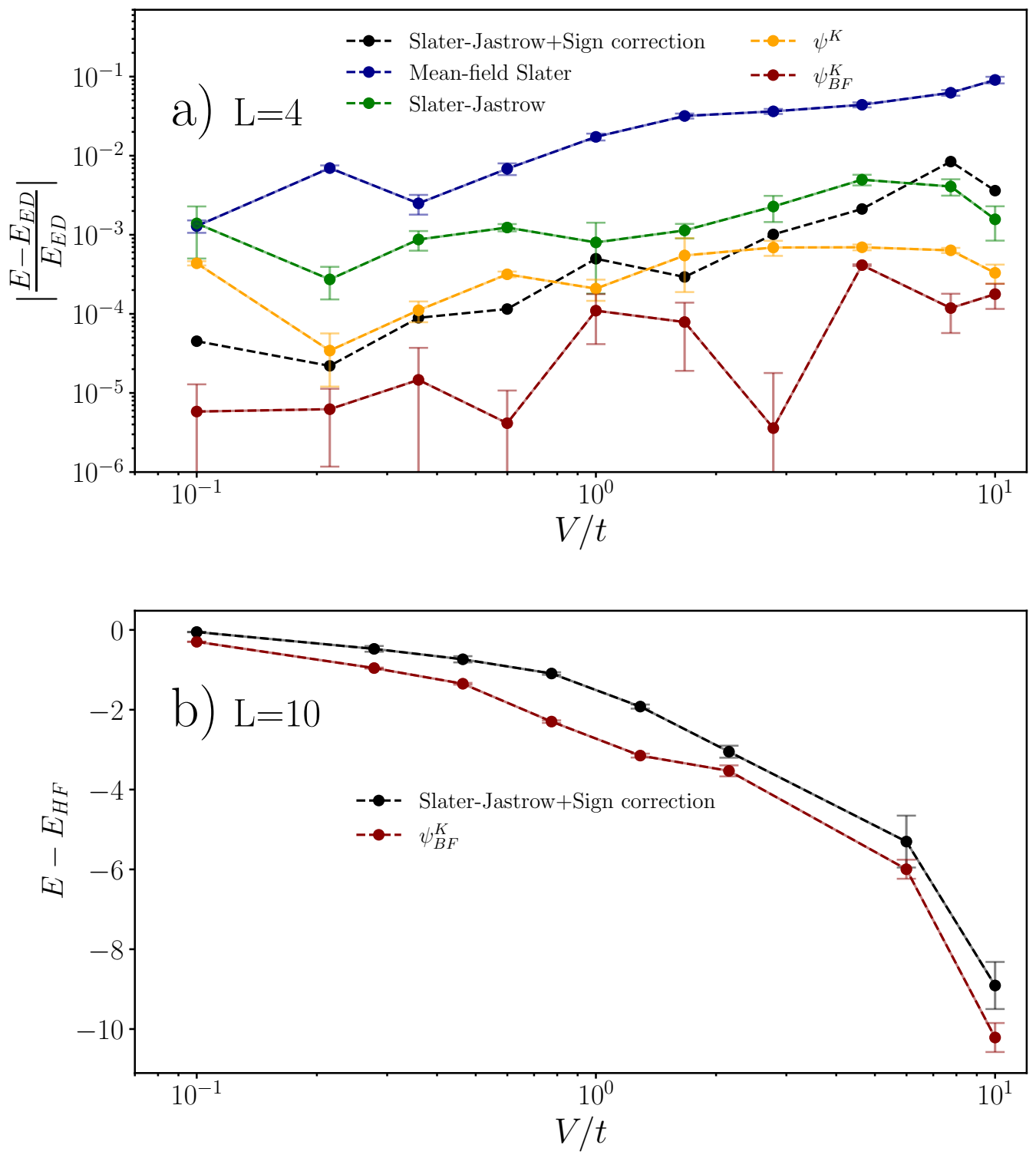

FIG. 1. **Comparison of ground-state energies of the $t$-$V$ model obtained with various wavefunction models.** We evaluate the performance of the symmetrized Neural Slater-Backflow-Jastrow ($\psi^K_{BF}$, red) against several other wavefunction models: symmetrized Slater-Jastrow without backflow ($\psi^K$, yellow), non-symmetrized mean-field Slater (blue), and non-symmetrized Slater-Jastrow (green). Additionally, we compare to the neural quantum state "Slater-Jastrow with an additional sign correction neural network" described in Ref. [22] (black). In panel **a)**, results are presented for a lattice size $L = 4$ with $\bar{n} = 0.44$, showing relative errors compared to exact diagonalization (ED) as a function of the interaction strengths $V/t$. Panel **b)** shows the deviation from Hartree-Fock (HF) energies, defined as $E - E_{HF}$, for a system with $L = 10$ and a filling fraction $\bar{n} = 0.44$. Error bars in both panels represent the corresponding standard deviations, where provided.

In Fig. 2 (a), we show the density-density correlation function as defined in Eq. 2 for $L = 10$ close to half-filling $\bar{n} = 0.44$ (see Supplementary Note II for results on the $8 \times 8$ system). When the interaction strength increases, the correlations start to oscillate more pronouncedly as a result of the increasingly ordered charge distribution. In the CO phase, the amplitudes of the oscillations decrease more gradually with distance compared to that of the metallic phase. For weak couplings, the correlations barely oscillate and fade as the graph distance $R$ increases.

To pinpoint the transition point in the thermodynamic limit, we use finite-size scaling. We study the structure factor $S(\pi, \pi)/N_s$ for various $V/t$. The critical transition point is found where the structure factor attains a finite value in the thermodynamic limit. In Fig. 2 (b) we plot the structure factor $S(\pi, \pi)/N_s$ as a function of the inverse system size $1/L$ (for $L = 4, 6, 8, 10$) and in Table I we report the extrapolated results in the thermodynamic limit.

We apply a least squares regression using the SciPy [74] library to fit the data and extrapolate $S(\pi, \pi)/N_s$ to the thermodynamic limit. In the metallic phase, the structure factor is expected to vanish for large system sizes, such that there exists a finite critical system size, $L_c$, beyond which the order parameter vanishes. This reflects the short-range nature of correlations in the metallic phase, which become negligible as the system size increases, causing the structure factor to diminish and ultimately disappear. In contrast, in the CO phase, $L_c$ tends to infinity, meaning the structure factor remains finite in the thermodynamic limit. The critical interaction strength, $V_c/t$, was identified at the point where the extrapolated $S_\infty$ transitions from the metallic phase to a positive finite value in the CO phase, signaling the onset of long-range order. Error estimates for $V_c/t$ were calculated using error propagation, based on the uncertainties in $S_\infty$ at these transition points.

Near half-filling, specifically for $\bar{n} = 0.44$, we estimate that the transition occurs at $V_c/t \simeq 1.14 \pm 0.04$, which is consistent with the value reported in Ref. [22]. While the order parameter from Eq. (3) of Ref. [22] explicitly assumes a symmetry-broken ground state, we enforce the symmetry and nevertheless find a transition point compatible with Ref. [22] using the density-density structure factor. We selected a density close to half-filling to observe the metallic to CO phase transition while avoiding the need for interpolating densities. This choice allowed us to explore conditions slightly doped away from half-filling, given that the phases at exact half-filling have already been extensively studied [22, 50, 59, 60, 62, 63].

### C. Excitations

To capture low-lying excitations, we carry out VMC optimizations across various momentum sectors $\mathbf{K} = (k_u, k_v)$, where $k_{u,v} \in \left\{0, \pm\frac{2\pi m}{L_{u,v}} \,\middle|\, m = 1, 2, \ldots, \frac{L_{u,v}}{2}\right\}$.

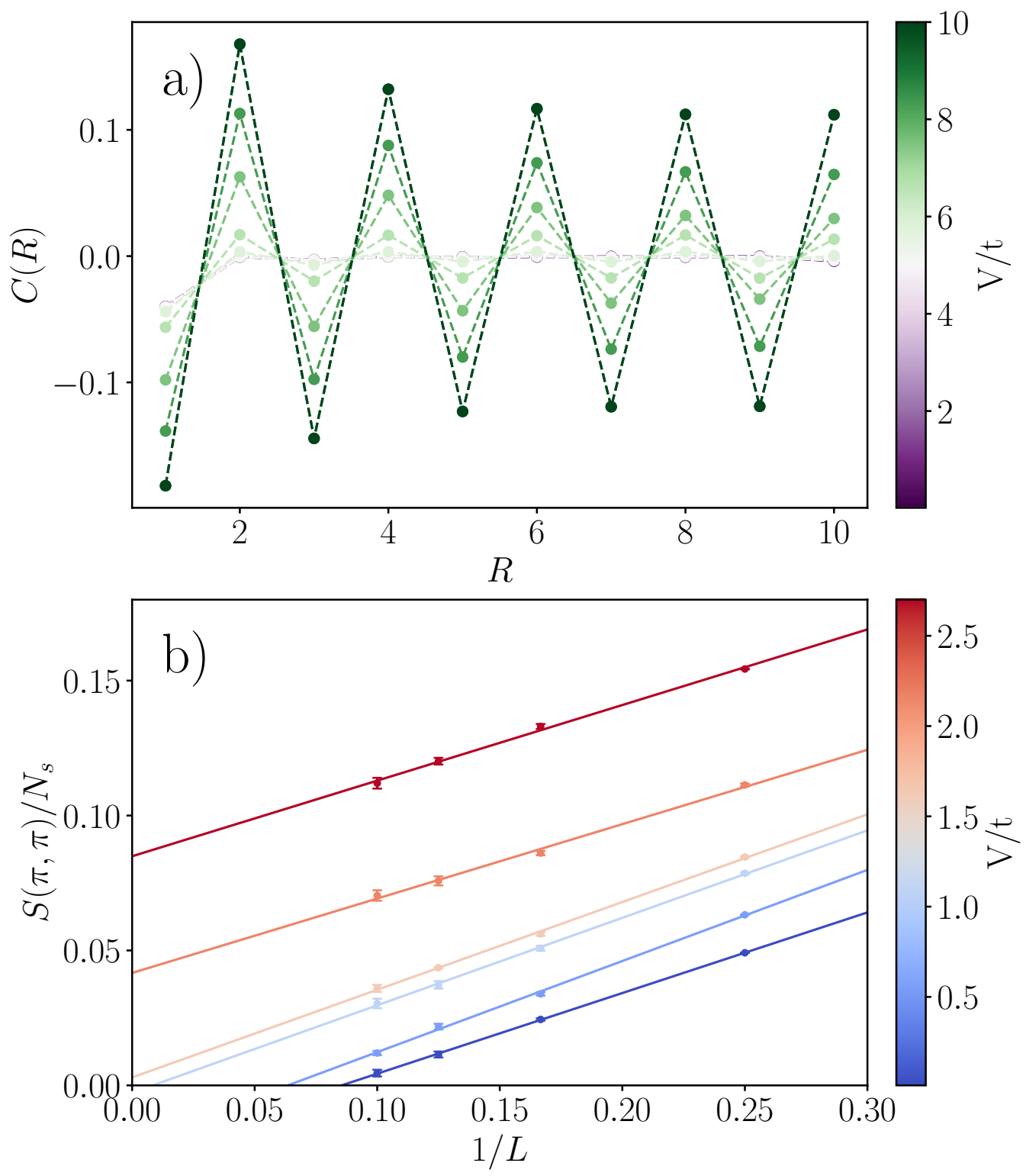

FIG. 2. **Two-point correlations and finite-size scaling of the structure factor. a)** The two-point density-density correlation functions as a function of graph distance $R$ for different interaction strengths $V/t$ for a system size of $L = 10$ and particle density $\bar{n} = 0.44$. **b)** Finite-size scaling of the structure factor $S(\pi, \pi)/N_s$ versus different system sizes $1/L$ (where $L = 4, 6, 8, 10$) with $\bar{n} = 0.44$ for different values of $V/t$ close to the estimated critical point $V_c/t \simeq 1.14 \pm 0.04$. Error bars represent the corresponding standard deviations.

| $V/t$ | $S_\infty$ |
|---|---|
| 0.1 | 0.0 |
| 0.6 | 0.0 |
| 1.0 | 0.0 |
| 1.3 | $0.003 \pm 0.001$ |
| 1.6 | $0.042 \pm 0.002$ |
| 2.7 | $0.085 \pm 0.002$ |

TABLE I. **Thermodynamic limit of the structure factor.** Structure factor extrapolations to the thermodynamic limit (taken from Figure 2 (b)) as a function of interaction $V/t$. Values in the metallic phase have been adjusted to reflect physical zeros for consistency.

Here $L_u$ and $L_v$ are the side lengths of the two-dimensional lattice in the $\mathbf{e}_u$ and $\mathbf{e}_v$ directions. In Fig. 3, we represent a single quadrant within the corresponding first Brillouin zone with conventional symbols to represent high-symmetry points [75].

We first benchmark the performance of our symmetrized Neural Slater-Backflow-Jastrow variational ansatz ($\psi_{BF}^K$) and a symmetrized Slater-Jastrow ($\psi^K$) (top) model for a system size of $L = 4$ and $\bar{n} = 0.31$ in Fig. 4. We observe that also in symmetry sectors different from the ground state, the symmetric backflow transformation reduces the relative energy error. We show the lowest energy state in each sector and compare it with results obtained from ED. In the lower panel of Fig. 4, we calculate the corresponding relative errors with respect to the ED energies. For our backflowed ansatz ($\psi_{BF}^K$) these lie between $10^{-6} - 10^{-3}$ for all sectors. In Supplementary Note IV, we also include additional simulation data of the open shell $L = 4$ system (see Fig. S3).

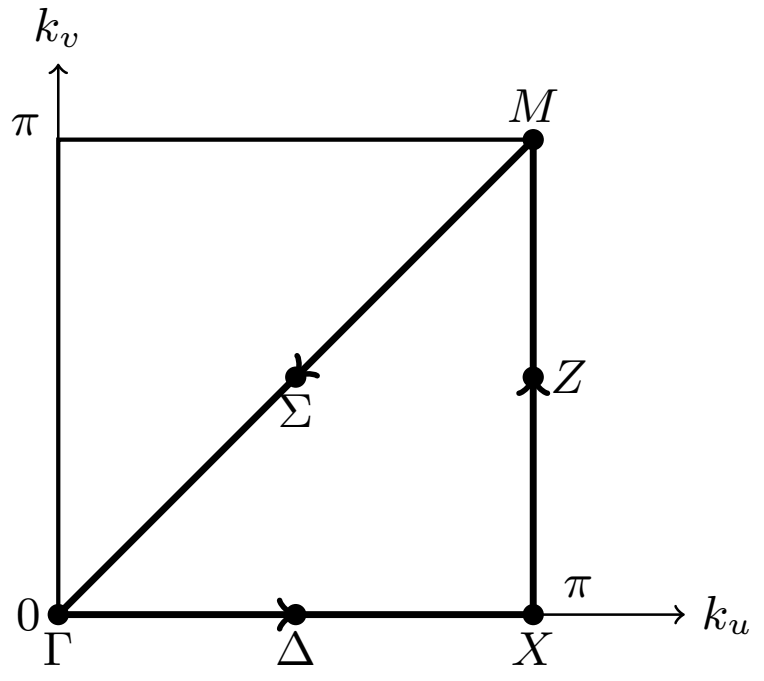

FIG. 3. **High-symmetry points in the Brillouin zone.** Single quadrant of the first Brillouin zone of the two-dimensional square lattice, highlighting the high-symmetry points $(k_u, k_v)$.

Prior studies have documented a range of coexistence phenomena for large and finite $V/t$ away from half-filling, transitioning from phase separation to potential stripe and checkerboard coexistence. In Ref. [76], the $t$-$V$ model on a square lattice with nearest-neighbor repulsive interactions was studied using mean-field theory for small system sizes. The authors observed a second-order phase transition from the Fermi liquid to the $(\pi, \pi)$ charge density wave state. At stronger repulsion, charge density waves coexisted at different momentum sectors when doped away from half-filling. In Ref. [53] and the subsequent study in Ref. [54], exact diagonalization was used to study small 2D systems. They found that at high repulsion and around quarter-filling densities, doped holes formed stable charged stripes acting as anti-phase walls [77], which are stable against phase separation in fermionic systems.

We extend our analysis to larger system sizes $L = 8$ with $\bar{n} = 0.39$ and $L = 10$ with $\bar{n} = 0.41$ to reduce finite-size effects. We simulate each system size with distinct particle densities to ensure they remain within the closed-shell domain, avoiding ground-state degeneracies in the non-interacting limit. We confirm the persistent non-degenerate ground state at the $\Gamma = (0, 0)$ point (see Fig. 5 for $L = 8$ and Fig. S2 in Supplementary Note III for $L = 10$). By comparing our symmetrized Neural Slater-Backflow-Jastrow ($\psi_{BF}^K$) ansatz with a symmetrized Slater-Jastrow ($\psi^K$) without backflow, we observe improved energy levels, even for low-lying excited

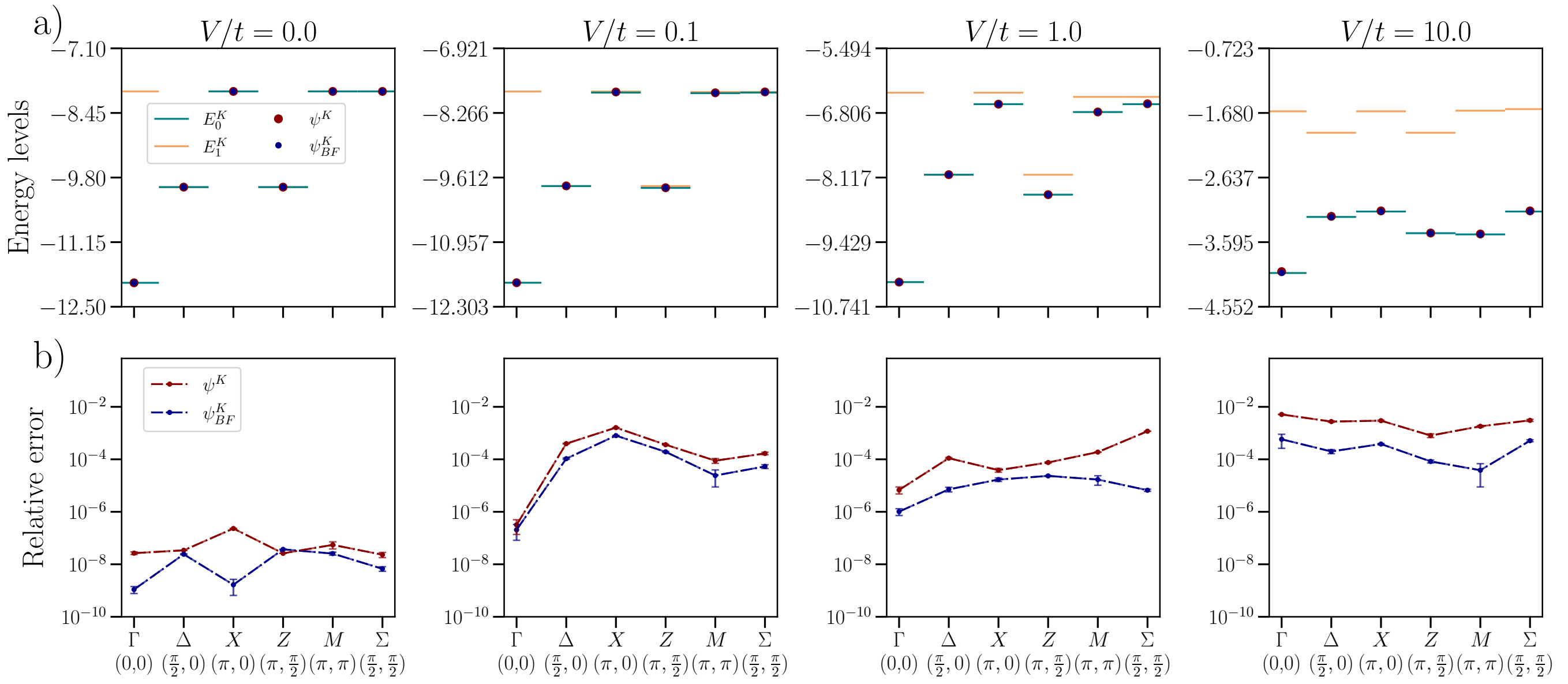

FIG. 4. **Energy comparison and relative errors for closed-shell** $L = 4$**. a)** The lowest ($E_0^K$, teal horizontal lines) and second-lowest energies ($E_1^K$, yellow horizontal lines) for the designated $\mathbf{K}$ sector using exact diagonalization (ED), compared to the lowest energies in each sector obtained with our variational ansatzes: symmetrized Neural Slater-Backflow-Jastrow ($\psi_{BF}^K$, blue) and symmetrized Slater-Jastrow without backflow ($\psi^K$, red). **b)** The relative energy errors associated with the ansatzes with respect to the lowest energy in each sector. These results are for a closed-shell system with lattice size $L = 4$ and a density of $\bar{n} = 0.31$. Error bars represent the corresponding standard deviations.

states, with the inclusion of the backflow correction term.

In Fig. 6 we depict the gap between the ground state $\Gamma$ and the excited energy levels in different sectors for different $V/t$ for both system sizes. We define the gap as the difference between the lowest energies in each sector relative to the lowest energy in the $\Gamma$ sector: $\Delta K = |E_0[\Gamma] - E_0[K]|$, where $K$ corresponds to the symbols of different sectors (here $K = M, X$) and $E_0[K]$ is the lowest energy in given sector. Notably, for strong interactions, we consistently observe a smaller gap for $\Delta M$ than for $\Delta X$. We include the gap $\Delta K_\infty$ for the $V/t = \infty$ value in the plot for both system sizes. This demonstrates a collapse in the interactions at infinite strength, indicating a charge-ordered phase where the electrons are fully localized. Furthermore, the gap between the lowest energy states in the $M$ and $X$ sectors appears largest in the intermediate coupling regime.

### D. V-score

We now aim to generalize the assessment of the performance of our model. Since ED becomes intractable for larger system sizes, we rely on the recently introduced V-score as a guiding metric [78], which can be computed using the variational energy and its variance. The V-score is dimensionless and invariant under energy shifts by construction. It is defined as [78]:

$$\text{V-score} = \frac{N \,\mathrm{Var}\, E}{(E - E_\infty)^2}, \tag{4}$$

where $N = N_f$ the number of degrees of freedom, $\mathrm{Var}\, E$ is the variance, $E$ is the variational energy and for the $t$-$V$ model

$$E_\infty = \frac{V|\mathcal{E}|N_f\,(N_f - 1)}{N_s\,(N_s - 1)}, \tag{5}$$

where $V$ is the interaction strength and $|\mathcal{E}|$ is the number of nearest neighbor bonds. The constant $E_\infty$ is used to compensate for global shifts in the energy, depending on the definition of the Hamiltonian.

The V-score serves as a valuable tool for discerning which Hamiltonians and regimes pose challenges for arbitrary classical variational techniques, even when we lack prior knowledge about the precise exact solution. Its practicality lies in its ability to quantify the accuracy of a particular method independently, without the need for direct comparisons with other methods. The intuition behind the V-score is that the energy variance, which becomes zero for an exact ground state, serves as a direct measure of how close a variational state is to the true ground state. This allows us to infer the accuracy of a method based on the variance alone. In particular, this metric enables us to draw comparisons between the accuracy obtained with our method on the given Hamiltonian, compared to other commonly studied condensed matter Hamiltonians, including spin Hamiltonians.

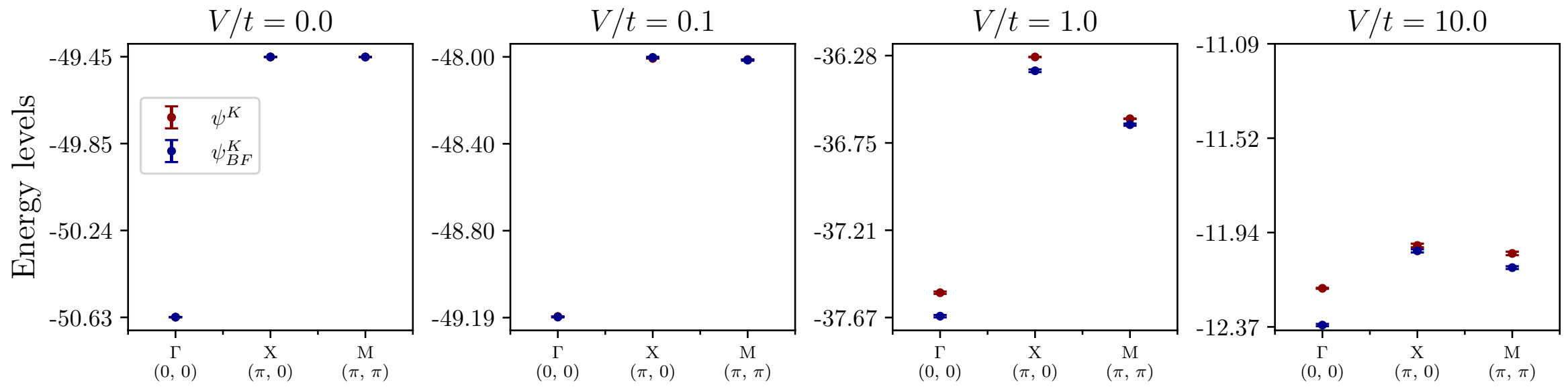

FIG. 5. **Lowest excitation energies for $L = 8$ closed-shell system.** Lowest excitation energies in different **K** sectors for a lattice of size $L = 8$, corresponding to a closed-shell configuration with particle density $\bar{n} = 0.39$, evaluated across various interaction strengths $V/t$. Results are shown for symmetrized Neural Slater-Backflow-Jastrow ($\psi^K_{BF}$, blue) and symmetrized Slater-Jastrow without backflow ($\psi^K$, red). Error bars represent the corresponding standard deviations.

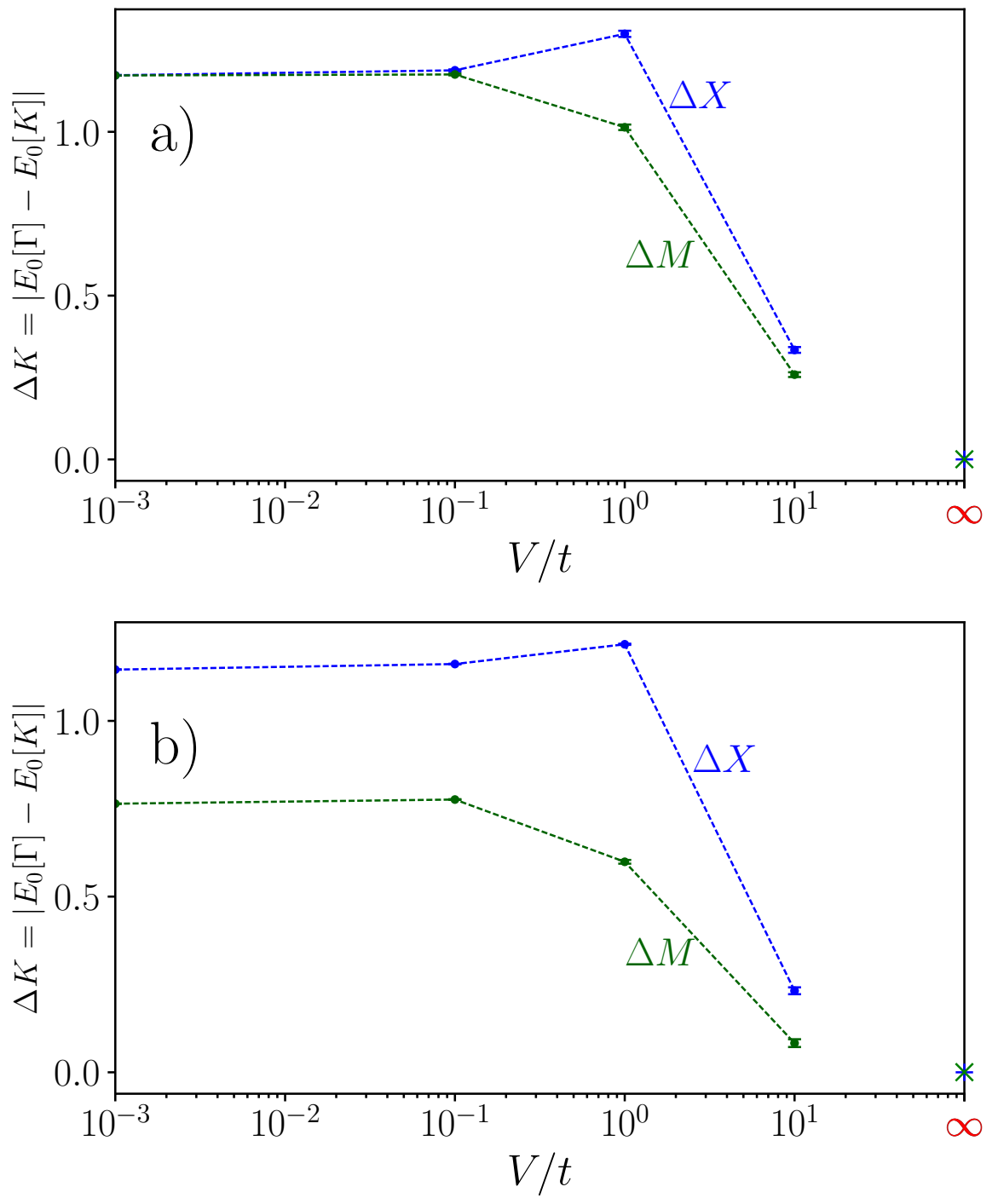

FIG. 6. **Energy gaps and excitation energies for closed-shell systems.** Energy gaps $\Delta K$ between the ground state (located at the $\Gamma$ point) and excitation energies ($K = M$, green; $K = X$, blue) for **a)** $L = 8$ lattice with filling factor $\bar{n} = 0.39$ and **b)** $L = 10$ lattice with filling factor $\bar{n} = 0.41$, both being closed-shell systems with no ground state degeneracy. We include $\Delta K_\infty$ for interaction strength $V/t = \infty$ in both figures, showing a collapse to a fully localized, charge-ordered phase at infinite repulsion. Error bars represent the corresponding standard deviations.

In Fig. 7, we present the ground state V-scores for different ansatzes, including the symmetrized Neural Slater-Backflow-Jastrow ($\psi^K_{BF}$), symmetrized Slater-Jastrow ($\psi^K$) and Hartree-Fock (HF) ansatz, for system sizes $L = 4$ with $\bar{n} = 0.31$, and $L = 10$ with $\bar{n} = 0.41$. The data clearly illustrate a strong dependence of V-scores on the specific interaction regime under investigation. In particular, as the V score values increase, it becomes increasingly challenging to achieve accurate solutions using demanding variational algorithms, implying a greater level of difficulty in solving these systems accurately. Despite these challenges, we observe that the backflow ansatz consistently exhibits lower V-scores compared to other methods, indicating its more accurate performance.

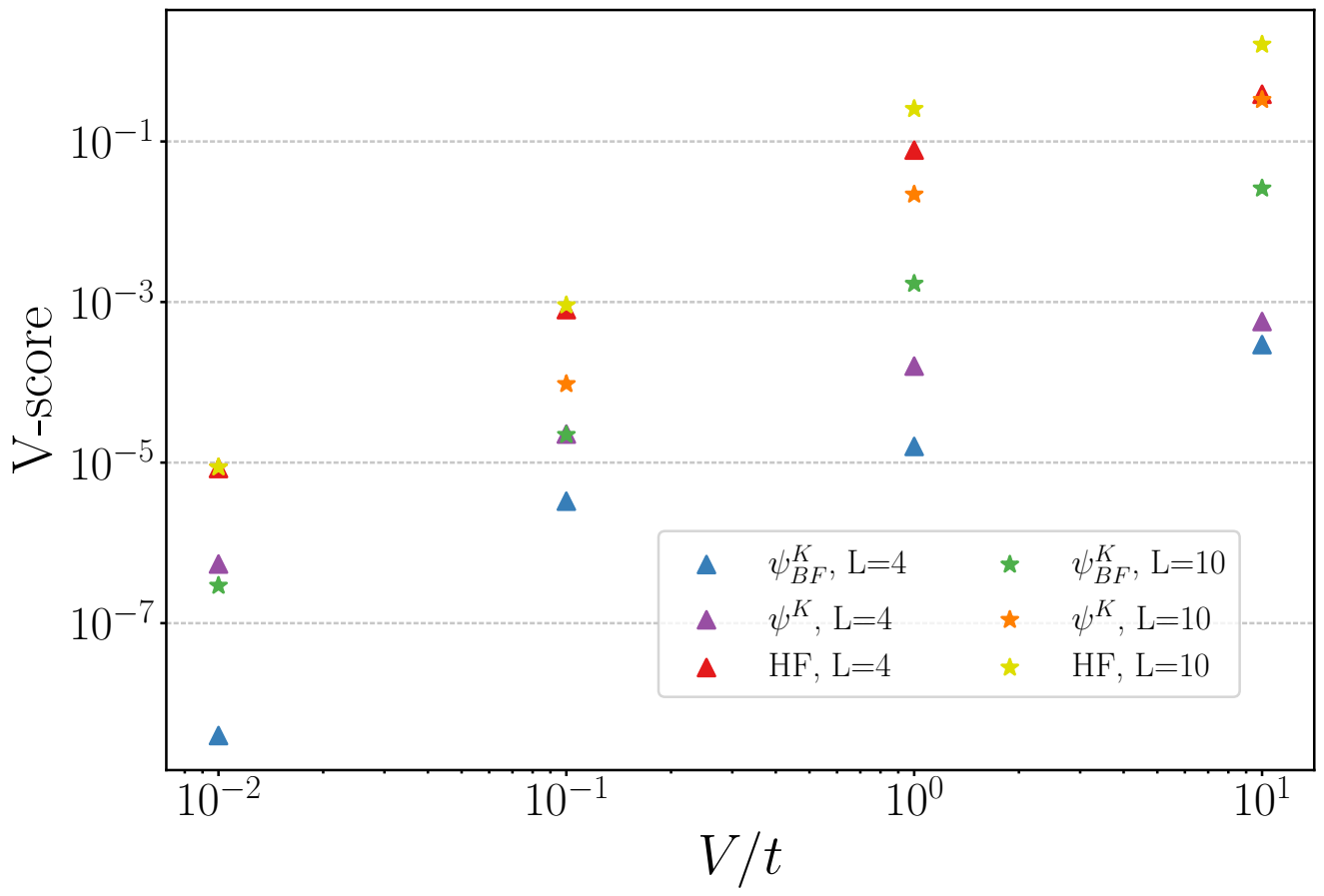

FIG. 7. **V-score comparison for different variational ansatzes.** Ground state V-scores as a function of interaction strength $V/t$ are compared for different ansatzes: $\psi^K_{BF}$ (symmetrized backflow Slater-Jastrow, blue/green), $\psi^K$ (symmetrized Slater-Jastrow, violet/orange), and HF (Hartree-Fock, red/yellow), shown for system sizes $L = 4$ with particle density $\bar{n} = 0.31$ (triangles) and $L = 10$ with $\bar{n} = 0.41$ (stars).

Next, we analyze the V-scores of our backflow ansatz for ground states and excitations, as depicted in Fig. 8, across various scenarios and closed-shell system sizes ($L = 4$ with $\bar{n} = 0.31$, $L = 8$ with $\bar{n} = 0.39$, and $L = 10$ with $\bar{n} = 0.41$). We compute the scores across a range of interactions $V/t$. We observe that we obtain similar V-scores for excited states as for ground states at all system sizes and interaction strengths. This indicates that our results for excited states are highly accurate, even in the large $V/t$ regime.

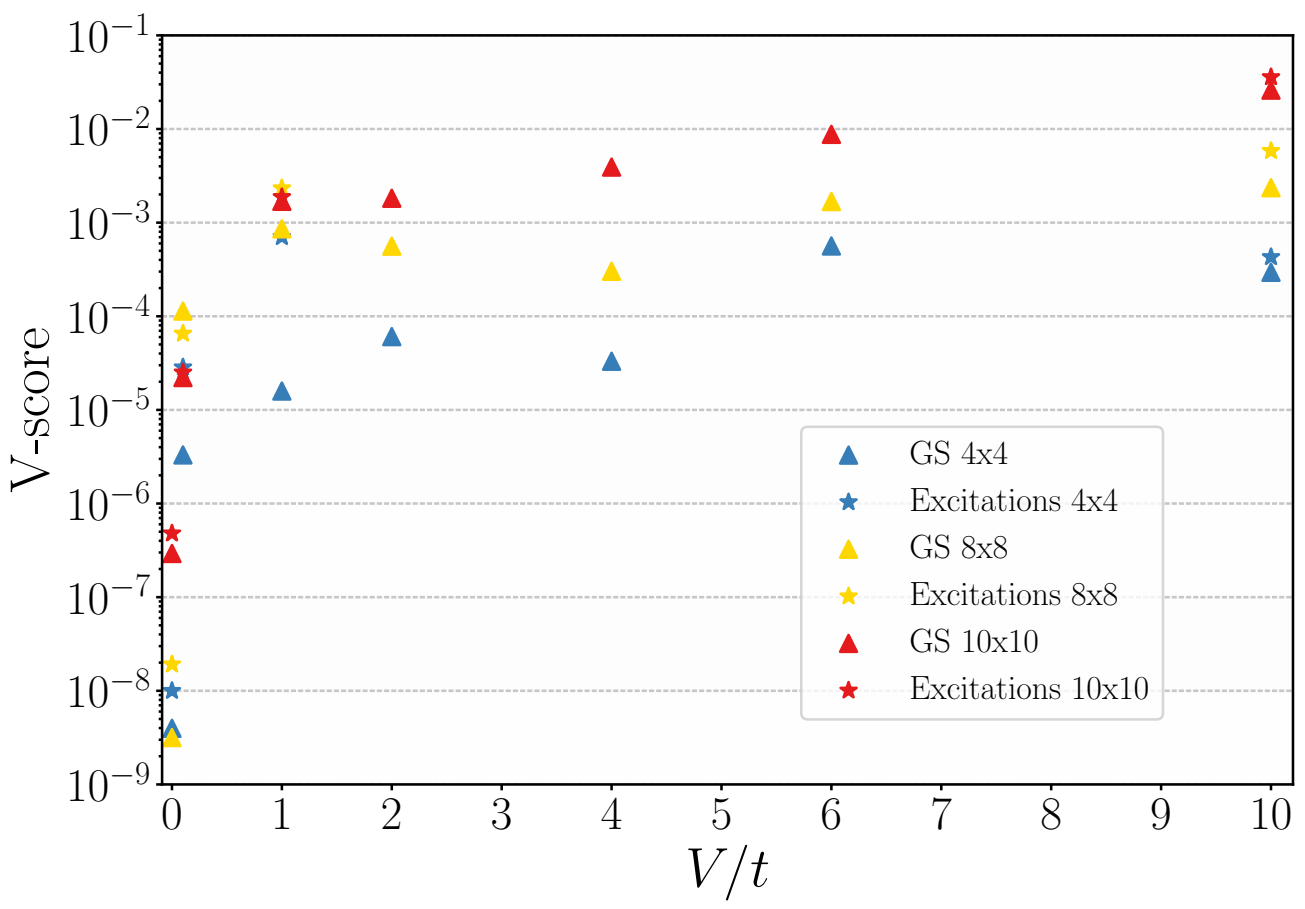

FIG. 8. **V-score comparison across system sizes and states.** V-score for a symmetrized Neural Slater-Backflow-Jastrow as a function of interaction strengths $V/t$ for ground (GS, triangles) and excited states (stars). Data are presented for different system sizes $L$ and particle densities $\bar{n}$: $L = 4$ with $\bar{n} = 0.31$ (blue), $L = 8$ with $\bar{n} = 0.39$ (yellow), and $L = 10$ with $\bar{n} = 0.41$ (red). Note that excitation data are available for $V/t$ values in the set {0.0, 0.1, 1.0, 10.0}, while ground-state data cover a broader range, including {0.0, 0.1, 1.0, 2.0, 4.0, 6.0, 10.0}.

## III. DISCUSSION

In this work, we introduced an approach to studying the low-energy excitation spectrum of fermionic Hamiltonians. By introducing symmetry-aware neural backflow transformations, we show that we can target the eigenstates of fermionic Hamiltonians with high accuracy. As a benchmark comparison, we show that this approach also yields significantly more accurate ground-state energies than other state-of-the-art variational Monte Carlo approaches.

In particular, we introduce equivariance conditions for the backflow transformations that lead to an efficient symmetry projection. We show that convolutional neural networks yield a powerful parametrization for our symmetry-aware backflow transformations that fulfill the equivariance conditions for both translation and particle-permutation symmetry. This key contribution enables us to efficiently access excited states by varying the total momentum $\mathbf{K}$ in the quantum number projection. Furthermore, we have showcased the utility of our approach in identifying phase-transitions on the $t$-$V$ model at system sizes far beyond what is reachable with exact diagonalization. To this end, we computed correlation functions and structure factors, resulting in the pinpointing of the critical point at $V_c/t = 1.14$. We also computed the V-score to quantify the variational accuracy of our proposed ansatz for different interaction regimes, system sizes, and excitations. Previous analysis based on the V-score has highlighted the challenging nature of targeting fermionic eigenstates. Our observations indicate that the symmetry-aware backflow ansatz yields accurate ground states and performs favorably compared to other state-of-the-art methods over the full interaction regime, including when strong correlations occur. Additionally, we find that our method yields accurate approximations to the low-energy eigenstates with a given $K$ momentum.

Future extensions of our approach include generalizing it by including additional symmetries, such as rotational and reflection symmetries. This will involve using more general group-convolutional kernels, as in group-convolutional neural networks (GCNN) [79–81]. This would require addressing the increased computational demands as the number of symmetry elements grows. We focused on the nearest neighbor $t$-$V$ model, but an extension is to consider Hamiltonians where spin-degrees of freedom become relevant as well (such as the Fermi-Hubbard model), or where interactions beyond nearest neighbor terms become relevant.

## IV. METHODS

### A. Fermions on the lattice

Consider a system of fermions that reside on a lattice represented by an undirected graph $\mathcal{G} = (\mathcal{V}, \mathcal{E})$, with a set of vertices denoted $\mathcal{V}$ and undirected edges as $\mathcal{E}$. Each lattice site is labeled $i \in \mathcal{V}$, and the total number of sites is $N_s = |\mathcal{V}|$. To each vertex $i$ we associate a position vector $\mathbf{r}_i$. The total number of fermions is conserved and will be fixed at $N_f \leq N_s$, and the particle density is defined as $\bar{n} = \frac{N_f}{N_s}$. We introduce the creation and annihilation operators of the fermionic mode (or lattice site) $i$ as $\hat{c}_i^\dagger$ and $\hat{c}_i$, respectively, and do not consider the spin of the fermions. These operators respect the usual fermionic anticommutation relations. In addition, we also introduce the corresponding number operator $\hat{n}_i = \hat{c}_i^\dagger \hat{c}_i$.

It will prove useful to connect the two main formalisms for describing fermionic systems: first quantization, which labels the fermions, and second quantization where we consider the occupation number basis or a given orbital set (here the lattice sites). To establish the correspondence, we consider a canonical ordering of the lattice sites through their label assignment $i = \{1, ..., N_s\}$. The latter can be chosen arbitrarily, and in practice we choose a zigzag-like ordering in the case of the 2D lattice. This enables us to recover the particle positions in the first quantization framework from a given occupation number configuration in second quantization (see Ref. [22]). We introduce $x = (\mathbf{r}_{i_1}, \mathbf{r}_{i_2}, \ldots, \mathbf{r}_{i_{N_f}})$, where $i_p$ is the site index occupied by the $p^{\text{th}}$ electron, where the fermion number index is determined by the chosen canonical ordering. In other words, we have an ordered set of indices $i_1 < i_2 < .. < i_{N_f}$. Furthermore, we introduce the occupation number configuration $n = (n_1, \dots, n_{N_s}) \in \{0,1\}^{N_s}$. Hence, in this notation, the canonical ordering allows us to extract active or occupied lattice indices $\{i_p\}_{p=1,\dots,N_f}$, given the occupation numbers $n$. This establishes implicit mappings $x = x(n)$ and $n = n(x)$ (see Fig. 9).

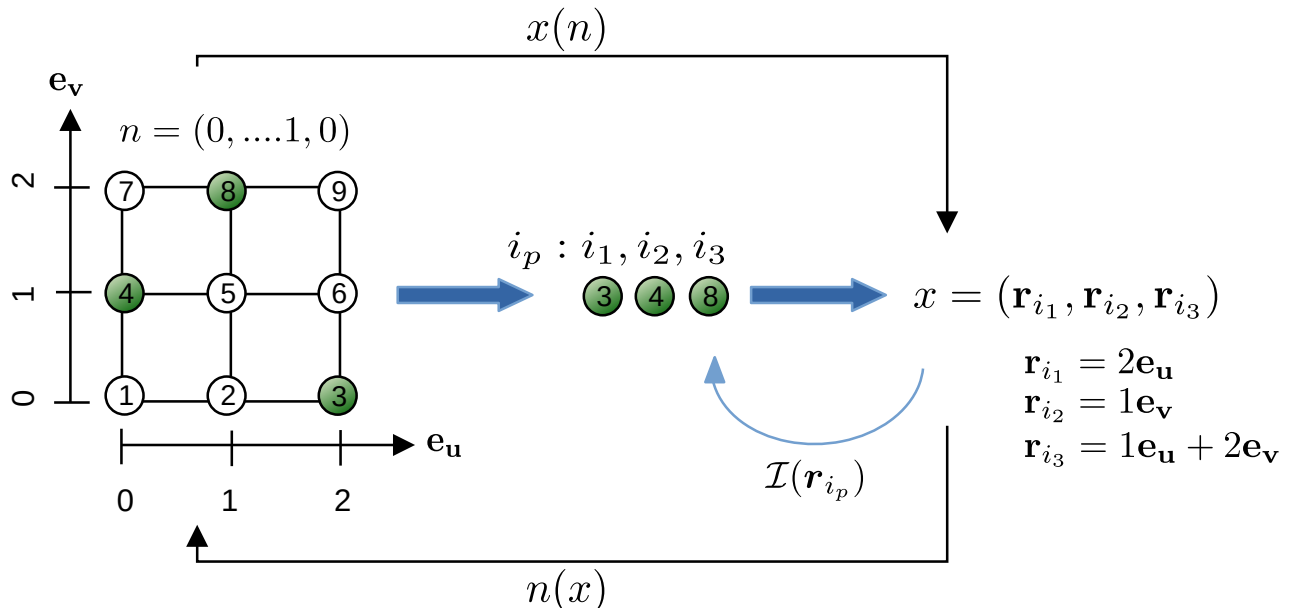

FIG. 9. **Implicit mappings between occupation $n(x)$ and position configurations $x(n)$.** This figure shows how the occupation configuration $n = (n_1, \dots, n_{N_s}) = (0,0,1,1,0,0,0,1,0)$ maps to the position vectors $x = (\mathbf{r}_{i_1}, \mathbf{r}_{i_2}, \dots, \mathbf{r}_{i_{N_f}})$, and vice versa, illustrating the relationship between occupied lattice sites and their indices respecting a canonical ordering.

## B. Wavefunction ansatz

### 1. Slater-Jastrow

Consider a set of $N_f$ single-particle mean-field (MF) orbitals $\{\phi_\mu(\mathbf{r})\}_{\mu=1,\dots,N_f}$ evaluated at position $\mathbf{r}$. For convenience, we introduce the matrix $M \in \mathbb{C}^{N_f \times N_s}$, with elements defined as

$$M_{\mu,i} = \phi_\mu(\mathbf{r}_i), \tag{6}$$

for all $N_s$ sites $i$. For a given set of particle positions $x$, we define the reduced matrix $\overline{M} \in \mathbb{C}^{N_f \times N_f}$ by selecting the columns of $M$ corresponding to the occupied sites: $\overline{M} = M_{\mu,i_p}$, where $i_p$ are the lattice sites that are occupied by particle $p \in \{1, .., N_f\}$.

The mean-field Slater determinant can be dressed with a Jastrow factor that introduces two-body correlations, and we obtain:

$$\psi(n) = \det \overline{M} \cdot e^{J(n)}. \tag{7}$$

The two-body Jastrow factor is defined as [29, 82]

$$J(n) = \frac{1}{2} \sum_{ij} n_i W_{d(ij)} n_j, \tag{8}$$

where the subscript $d(ij)$ of the complex variational parameters $W$ denotes the distance between site $i$ and $j$.

### 2. Neural Backflow Transformation

The Slater-Jastrow wavefunction ansatz can be further improved by including backflow transformations, thereby significantly increasing the expressiveness of the model. We use the neural backflow transformation that effectively promotes the single-particle orbitals to many-body orbitals [26, 27, 39]. Therefore, we introduce the backflow function $F$ that produces a configuration-dependent orbital matrix $F \in \mathbb{C}^{N_f \times N_s}$. We then adapt the orbital matrix as

$$M_{\mu,i} \to B_{\mu,i}(n) = M_{\mu i} \circ F_{\mu i}(n), \tag{9}$$

where $\circ$ corresponds to an element-wise product between the matrices $M$ and $F$. The corresponding reduced matrix of $B$ is $\bar{B}$, and is obtained similarly as for $\bar{M}$. The Neural Slater-Backflow-Jastrow ansatz is then defined as

$$\psi_{BF}(n) = \det \overline{B}(n) \cdot e^{J(n)}. \tag{10}$$

Below, we will describe the properties of the backflow function $F$ and introduce a neural parametrization thereof.

## C. Symmetries and Excitations

We consider an electronic Hamiltonian on a lattice that commutes with the elements of a symmetry group $G$, such as total spin, the total momentum, and geometrical symmetries such as rotation. The eigenstates of the many-body Hamiltonian can be classified with the symmetry sectors of $G$. We restrict the NQS ansatz to a given symmetry sector labeled by $I$ through a quantum-number projection

$$\psi^I(n) = \sum_{g \in G} \chi_g^{I*} \psi\left(\hat{g}^{-1} n\right), \tag{11}$$

and $\chi_g^I$ is the character corresponding to the irreducible representation (irrep) $I$ and group element $g$. To make the notation more explicit, consider a translation operator denoted by $\hat{g} = \hat{T}_{\boldsymbol{\tau}}$, where $\boldsymbol{\tau}$ is the corresponding translation vector. The effect of the operator on a configuration $n$ is to permute the site indices $(1, \dots, N_s) \to (\tau_1, \cdots, \tau_{N_s})$, i.e. $\hat{T}_{\boldsymbol{\tau}} |n\rangle = |n_{\tau_1}, \cdots, n_{\tau_{N_s}}\rangle$ where in terms of the position map $n_{\tau_i} = n(\boldsymbol{r}_i - \boldsymbol{\tau})$. In this work we focus on the projected form

$$\psi^K(n) = \sum_{\boldsymbol{\tau}} e^{-i\boldsymbol{\tau}\cdot\mathbf{K}} \psi\left(\hat{T}_{\boldsymbol{\tau}}^{-1} n\right), \tag{12}$$

where $\mathbf{K}$ is the total momentum and the sum runs over all possible translation vectors. Due to the antisymmetric nature of the wavefunction, translating fermions can lead to a change in the sign of the wavefunction, depending on how their positions are permuted. This effect is naturally handled by the above-mentioned quantum number projection with corresponding characters, ensuring that the fully symmetrized wavefunction remains translationally invariant.

We use this approach to compute both the ground state and the low-lying excited states. The low-lying excitations are characterized by different momentum sectors, and their computation involves optimizing the wavefunction within the quantum number sectors distinct from the ground state [43, 45]. Instead of focusing on a single excited state individually, an alternative strategy involves adopting a multi-target approach, which has recently been introduced for continuous systems [83]. However, for cost-effectiveness and interpretation in terms of translation quantum numbers, we focus on the symmetry-projection method outlined above.

In a brute-force approach, the evaluation of the symmetrized wave function $\psi^I(n)$ for a configuration $n$ would require $|G|$ evaluations of the parametrized non-symmetric wave function $\psi$ in Eq. (11). In particular, for translation symmetry in Eq. (12) of a square lattice of size $L \times L$, this would require $|G| = L^2$ evaluations. The computational burden induced by the symmetrization procedure can therefore become significant for increasing system sizes. For this purpose, we introduce a set of *symmetry-aware neural backflow transformations* that are constructed as equivariant functions, requiring only a single evaluation of a neural network to produce *all* $\psi(\hat{g}^{-1}n)$ (i.e. $\forall g \in G$) required to evaluate the projection in Eq. (11). This will allow us to reach larger system sizes, even when considering deep neural networks to represent the backflow transformation. In the next section, we discuss the requirements of this symmetry-aware backflow transformation and introduce a specific neural parametrization to fulfill the constraints.

### 1. *Equivariance Condition*

We will introduce backflow transformations that keep both the particle-permutation and lattice symmetries manifest, by introducing transformations that are equivariant under the respective groups. More concretely, when two fermions $p$ and $q$ are exchanged by the permutation operator $\hat{P}_{pq}$ or when a lattice-symmetry transformation $\hat{g}$ is applied to the lattice, the respective outputs of the neural backflow change accordingly:

$$F_{\mu,i_p}(\hat{P}_{pq}^{-1}n) \stackrel{!}{=} F_{\mu,\hat{P}_{pq}i_p}(n) = F_{\mu,i_q}(n), \tag{13}$$

$$F_{\mu,i_p}(\hat{g}^{-1}n) \stackrel{!}{=} F_{\mu,\hat{g}i_p}(n). \tag{14}$$

In the case of translations we enforce $F_{\mu,i_p}(\hat{T}_{\boldsymbol{\tau}}^{-1}n) = F_{\mu,\hat{T}_{\boldsymbol{\tau}}i_p}(n) = F_{\mu,\mathcal{I}(\boldsymbol{r}_{i_p}+\tau)}$, where $\mathcal{I}(\boldsymbol{r}_{i_p} + \boldsymbol{\tau})$ (defined in Fig. 9) denotes the index of the lattice site obtained by shifting $\boldsymbol{r}_{i_p}$ by the translation vector $\boldsymbol{\tau}$. In other words, lattice symmetries can be defined by their permutation of the lattice sites.

Our key objective is to preserve translation equivariance in the backflow transformation. Using this, we can construct a symmetrized Neural Slater-Backflow-Jastrow ansatz, given that the Jastrow correlation function is translation-invariant and the backflow is constructed as a neural network respecting the equivariance conditions in Eqs. (13) and (14). A natural candidate for a translation-equivariant neural network is a convolutional neural network operating on occupation configurations [79–81], which naturally exhibits these properties. In a CNN, spatial translations in the input lead to corresponding shifts in the output feature maps, effectively preserving spatial locality. The occupation configurations undergo multiple CNN-transformation layers, resulting in outputs of the same size $L^2$ as the input configuration. This preserves the structure needed to maintain equivariance, without averaging over symmetry group elements as done for invariance. We employ $N_f$ independent backflow transformations corresponding to the different orbitals $\mu$. From the outputs, we obtain the reduced matrix $\overline{B}$ in Eq. (10), by selecting the columns corresponding to the occupied sites from the resulting backflow matrix $F_{\mu,i}(n)$. We depict this procedure in Fig. 10 where we provide a concise visual representation of a CNN backflow satisfying the equivariance conditions. In summary, given our CNN backflow is equivariant: instead of evaluating the CNN for all elements of the translation symmetry group, we can extract all the required $\psi\left(\hat{g}^{-1}n\right)$ in Eq. (11) from the output of a single evaluation of the backflow CNN. This approach improves efficiency and reduces computational redundancy in handling symmetrical transformations. In Fig. 10, we illustrate this symmetry-averaging process under equivariance conditions. For additional information on the architecture and its adaptation to different system sizes, we refer to Supplementary Note I.

## ACKNOWLEDGMENTS

We thank Dian Wu and Javier Robledo Moreno for engaging discussions. We especially thank Javier Robledo Moreno for sharing the data of Ref. [22]. We express our gratitude to Yusuke Nomura for providing valuable insights regarding quantum phase transitions and their connection to the structure of excitations. This work was supported by the Swiss National Science Foundation under Grant No. 200021_200336, and Microsoft Research.

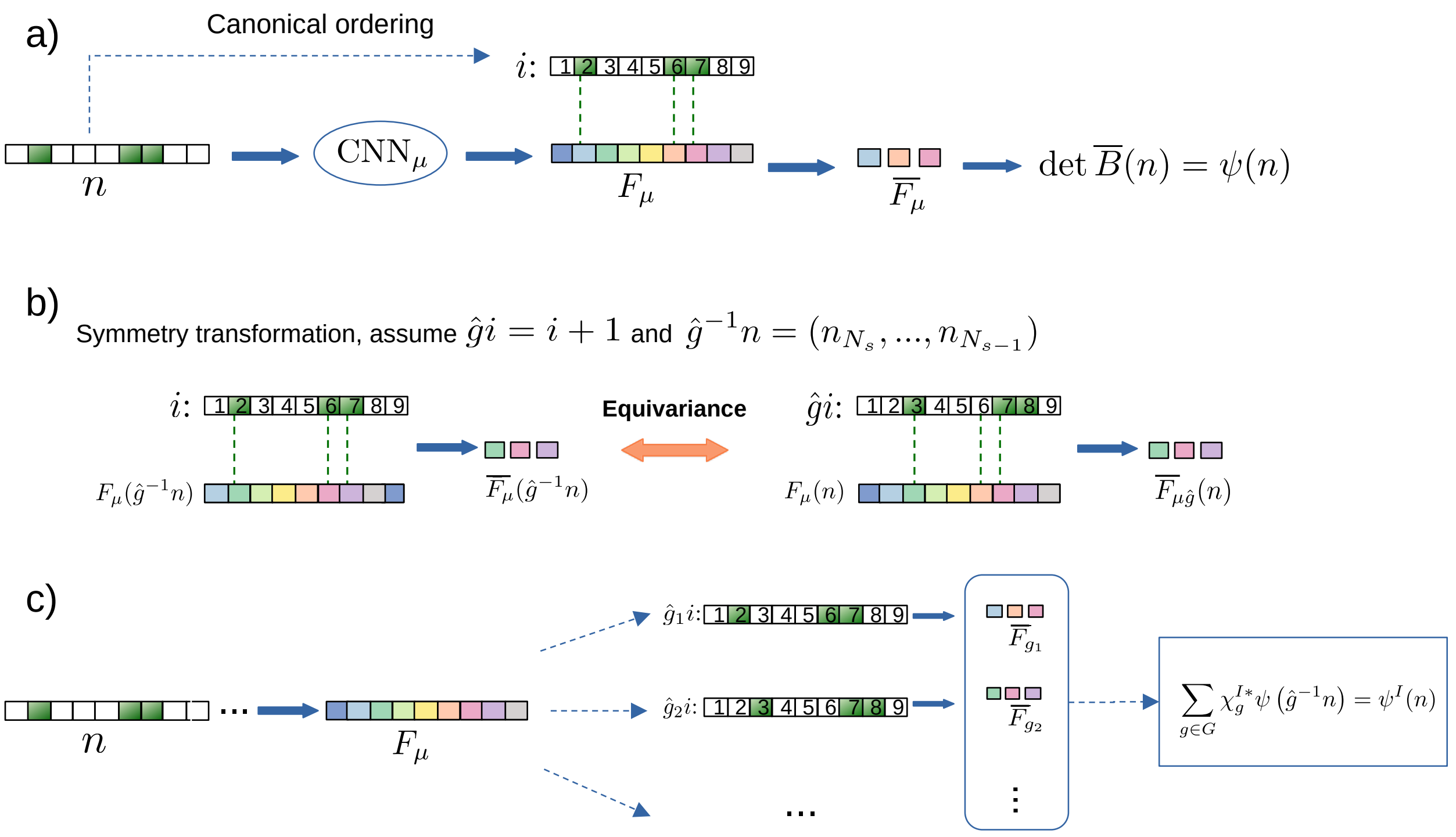

FIG. 10. **Representation of symmetry-aware backflow and quantum number projection.** Panel **a)** showcases how the backflow determinants are constructed. The neural network CNN$_\mu$ for a given orbital $\mu$ takes as input the configuration $n$, and produces a backflow vector output $F_\mu(n)$. The reduced matrix $\bar{F}_\mu$ is obtained by selecting the active indices $i$ from $F_\mu$. These active indices are linked to the occupied sites, represented by the dark green blocks in the input $n$, using the canonical ordering. Panel **b)** demonstrates the equivariance property of the backflow function. Applying a symmetry transformation to the input as $\hat{g}^{-1}n$ and then extracting the active indices, we get $\bar{F}_\mu(\hat{g}^{-1}n)$. This is equivalent to applying the symmetry transformation directly to the active indices represented by $\hat{g}i$. This results equivalently in the reduced matrix $\bar{F}_{\mu\hat{g}}(n)$. Panel **c)** represents the quantum number projection. Following a single CNN evaluation, we obtain the reduced matrices $\bar{F}_{\mu g}$, $\forall \hat{g} \in G$ by properly constructing the reduced matrix, without reevaluating the backflow CNN. Subsequently, we compute the symmetry-averaged wavefunction $\psi(\hat{g}^{-1}n)$.

## Supplementary Note I. NUMERICAL DETAILS AND BACKFLOW ARCHITECTURE

Given a set of occupation numbers, our symmetrized wavefunction ansatz is expressed as in Eq. 12 of the main text, where we use a translation-invariant Jastrow factor defined in Eq. 8 of the main text, with variational parameters $\theta_J = W_{d(ij)}$. The mean-field orbitals are captured by the matrix $M_{\mu,i}$ as defined in Eq. 6 of the main text, of variational parameters of size $N_f \times N_s$. The backflow corrections are constructed as a translational equivariant convolutional neural network with parameters $\theta_{BF}$.

The total set of variational parameters $\theta = (M_{\mu i}, \theta_{BF}, \theta_J)$ are all optimized simultaneously using Variational Monte Carlo and Stochastic Reconfiguration [84]. To accelerate the optimization convergence of our Neural Slater-Backflow-Jastrow ansatz, we initialize the orbital parameters with converged parameters of a Hartree-Fock optimization.

For smaller systems (i.e. $L = 4$) we use shallow backflow CNN networks with one hidden layer and 32 features. For larger and more challenging systems such as $L = 8, 10$ we construct a residual CNN with 5 convolutional layers and 16 features per filter. Residual networks make the training more stable [22, 81, 85]. To access low-lying energy states, we use the latter residual CNN architecture regardless of the system size. Our deep CNNs use complex-valued weights for the feature maps. Each layer, except the final one, uses the complex Rectified Linear Unit (ReLU) activation function [86]. The output layer produces the backflow functions necessary for our computations.

We optimize multiple CNN backflow transformation models in parallel, corresponding to the $N_f$ orbitals, ensuring efficient and scalable processing. For the optimization, we use a learning rate of 0.0075. The optimization is carried out using the SR preconditioner with a diagonal shift of 0.001, and we use 4096 Monte Carlo samples.

## Supplementary Note II. DENSITY-DENSITY CORRELATION FUNCTION

To complement the results for $L = 10$ shown in Fig. 2 (a), we also present the density-density correlation function, as defined in Eq. 2 of the main text, for a system size of $L = 8$ and $\bar{n} = 0.44$ in Fig. S1. Similarly to the $L = 10$ case, we observe that the correlations display more pronounced oscillations as $V/t$ increases, indicating the increasing orderliness of the charge distribution during the phase transition. This behavior suggests a stronger tendency towards charge ordering with increasing interaction strength. In the charge-ordered (CO) phase, the amplitude of the oscillations gradually diminishes with distance. Conversely, in the metallic phase, the oscillations diminish more quickly with distance, highlighting a less ordered charge distribution. For weak couplings, the correlations exhibit minimal oscillations and fade as the graph distance $R$ increases.

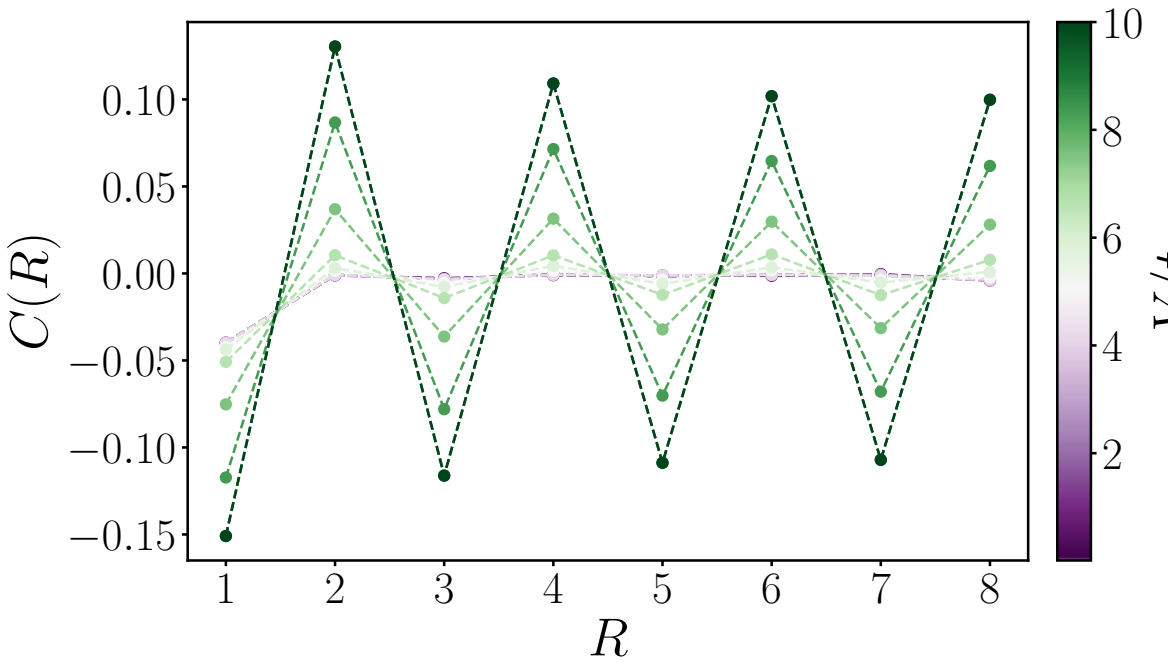

FIG. S1. **Two-point correlations for $L = 8$.** The two-point density-density correlation functions as a function of graph distance $R$ for different values of interaction strength $V/t$ for a system size of $L = 8$ and $\bar{n} = 0.44$ with ground state momentum $\mathbf{K} = (\frac{3\pi}{4}, 0)$.

## Supplementary Note III. EXCITATION SPECTRUM FOR CLOSED-SHELL $L = 10$ CASE

In Fig. S2, we present the energy spectrum for $L = 10$ and a closed-shell density $\bar{n} = 0.41$, across different symmetry sectors for varying $V/t$. Similarly to the $L = 8$ case discussed in the main text (Fig. 5), we observe that the ground state is located within the $\Gamma$ sector.

As in the previous case, we also compare our symmetrized Neural Slater-Backflow-Jastrow ($\psi^K_{BF}$) ansatz with a symmetrized Slater-Jastrow ($\psi^K$) ansatz without backflow, and observe improved energy levels, even for low-lying excited states, with the inclusion of the backflow correction term, especially for larger $V/t$ values.

## Supplementary Note IV. EXCITATION STRUCTURE FOR OPEN-SHELL $L = 4$ CASE

Moving away from closed shells induces changes in the structure of the excitation energies [87–89]. In Fig. S3, we benchmark our ansatz for the $L = 4$ system with $\bar{n} = 0.44$, which is not a closed-shell system. In the upper panel, we compute the lowest and second-lowest en-ergies using ED methods in each assigned **K** sector and compare the lowest energies in each sector to our VMC ansatzes. In the lower panel, we present the corresponding relative errors of the ansatzes. Our results show that the symmetric backflow improves the accuracy, demonstrating that our ansatz can also be effectively applied to non-closed-shell systems for computing excited states.

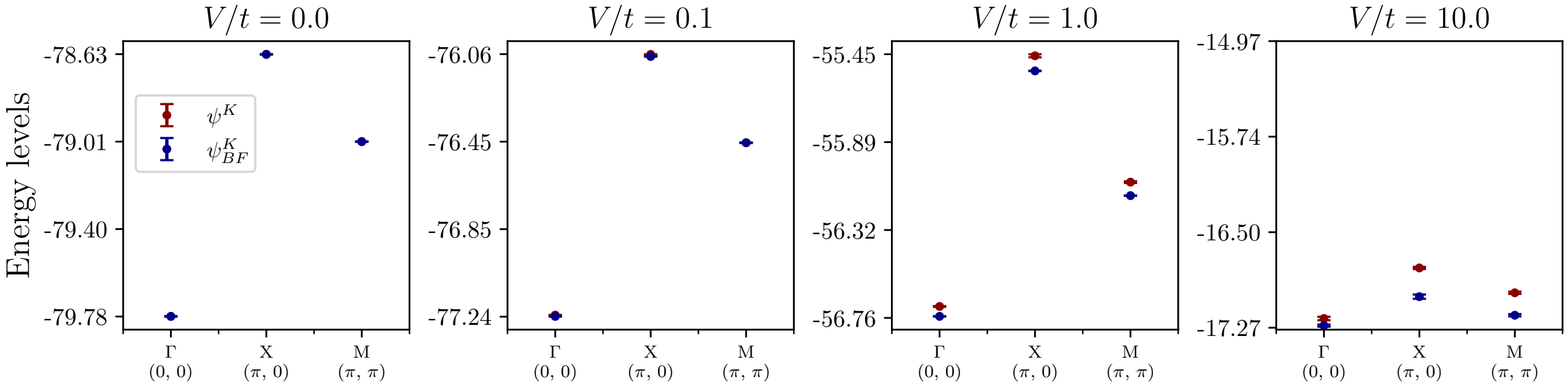

FIG. S2. **Lowest excitation energies for $L = 10$ closed-shell system.** Lowest excitation energies in different **K** sectors for an $L = 10$ closed-shell system with particle density $\bar{n} = 0.41$, evaluated across various interaction strengths $V/t$. Results compare different variational ansatzes: symmetrized Neural Slater-Backflow-Jastrow ($\psi^K_{BF}$, blue) and symmetrized Slater-Jastrow without backflow ($\psi^K$, red). Error bars represent the corresponding standard deviations.

## Supplementary Note V. EFFECT OF NETWORK SIZE AND SYMMETRY

In Fig. S4, we illustrate how the size of the neural network affects the accuracy of ground state energy accuracy for an $L = 8$ system at a fixed $V/t$ interaction. The converged VMC energy is shown as a function of different layer sizes $\ell$ for various backflow neural network architectures, ranging from an equivariant CNN to a simple feed-forward neural network. The results indicate that increasing network size enhances energy accuracy. Dashed lines represent the no-backflow ansatzes: mean-field Slater determinant, Slater-Jastrow, and symmetrized Slater-Jastrow ($\psi^K$). In contrast, solid lines show the inclusion of backflow: a simple feedforward NN backflow and a symmetrized Neural Slater-Backflow-Jastrow with an equivariant CNN ($\psi^K_{BF}$). Additionally, we include data comparing a symmetry-aware convolutional neural network backflow ansatz that is not equivariant, where symmetry is incorporated by averaging over the symmetry group elements ($\psi^K_{BF}$-invariant), also shown with solid lines. The inset shows the results without the mean-field Slater ansatz, providing a zoomed-in view of the plot. The equivariant CNN-based neural network achieves lower energies than its invariant counterpart and other ansatzes. Although the invariant neural network backflow averages over group elements, it may not leverage the symmetry as effectively as the equivariant backflow. The equivariant neural network backflow integrates symmetry directly into each orbital's backflow.

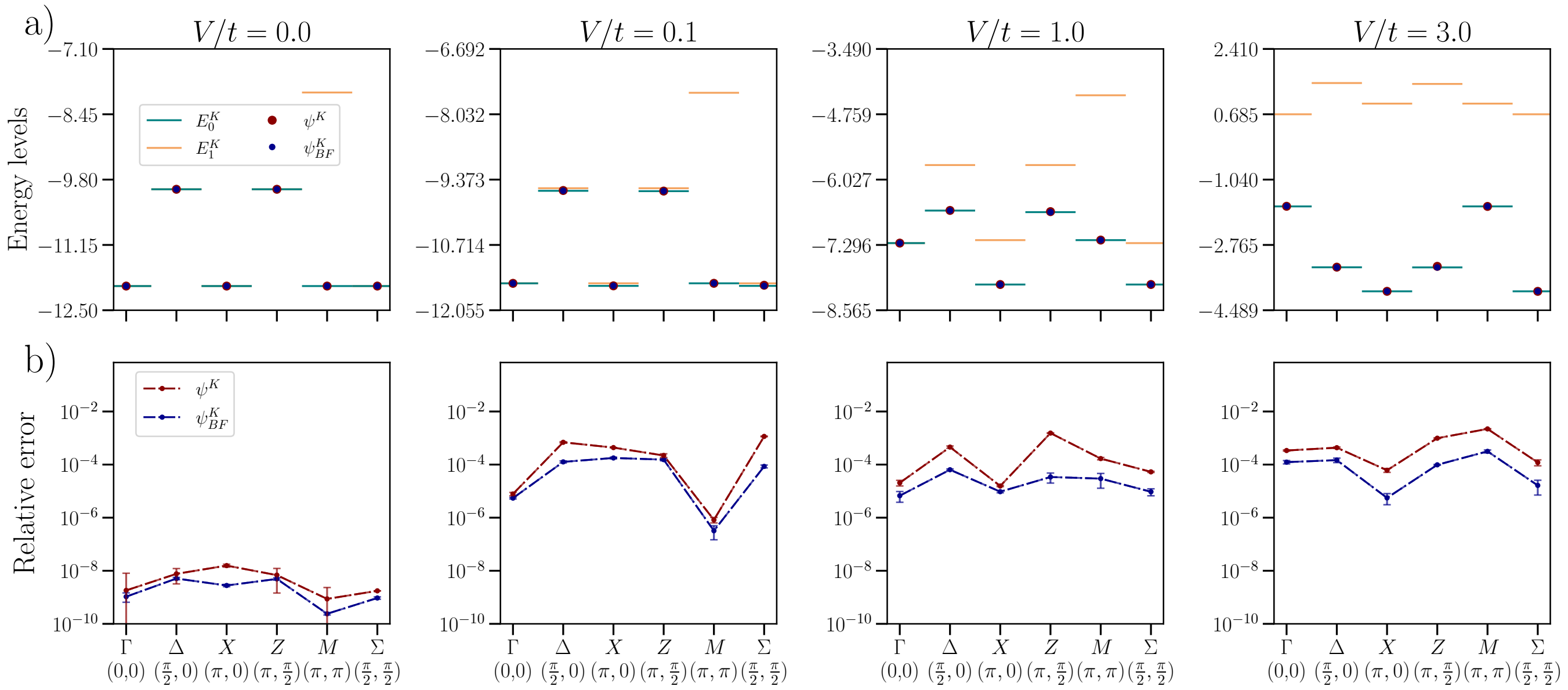

FIG. S3. **Energy comparison and relative errors for $L = 4$ open-shell system.** **a)** The lowest ($E_0^K$, teal horizontal lines) and second-lowest energies ($E_1^K$, yellow horizontal lines) for the designated $\mathbf{K}$ sector using exact diagonalization (ED), compared to the lowest energies in each sector obtained with our variational ansatzes: symmetrized Neural Slater-Backflow-Jastrow ($\psi^K_{BF}$, blue) and symmetrized Slater-Jastrow without backflow ($\psi^K$, red). **b):** The relative energy errors of the ansatzes with respect to the lowest energy in each sector are shown as dashed lines. These results correspond to a system with lattice size $L = 4$ and density $\bar{n} = 0.44$, which is not a closed-shell system. Error bars represent the corresponding standard deviations.

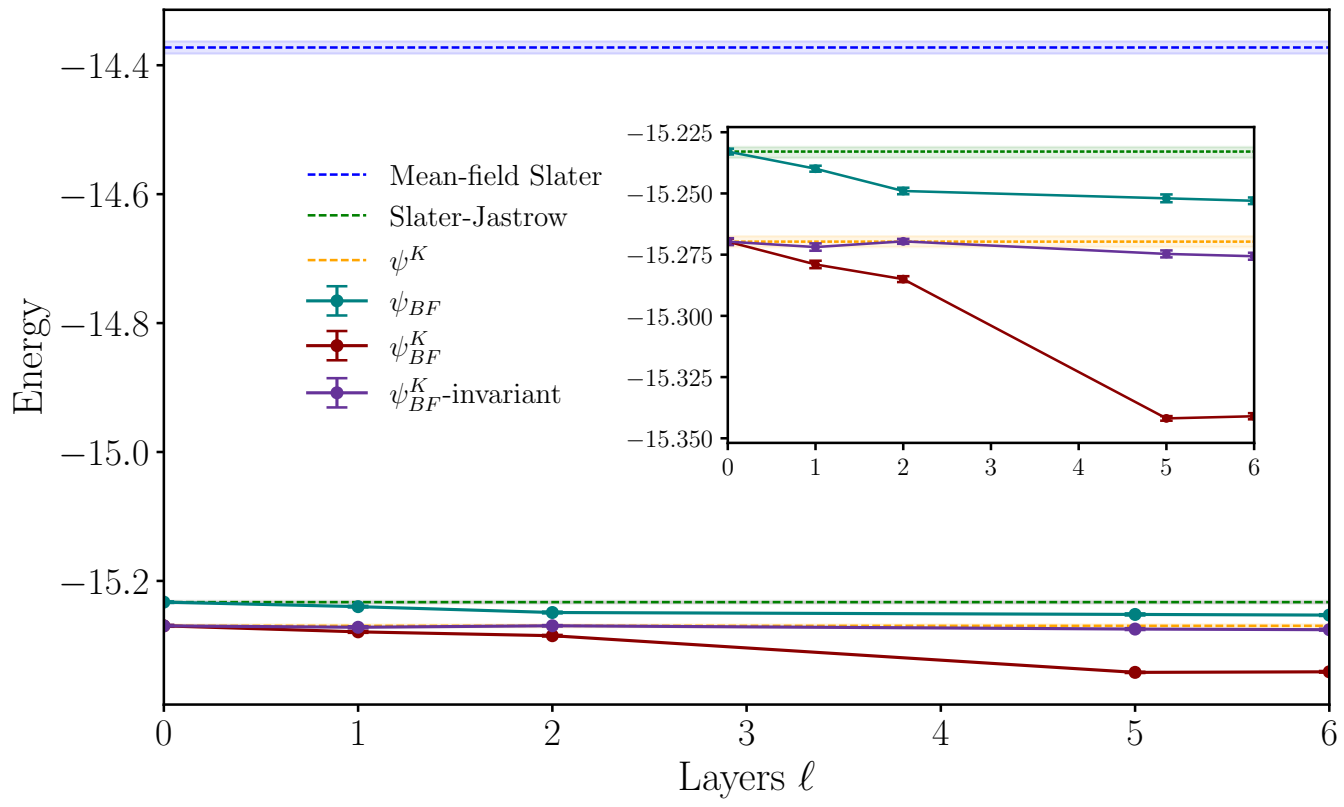

FIG. S4. **Impact of neural network size on energy accuracy.** Effect of neural network size on ground state energy accuracy for a system of size $L = 8$, with density $\bar{n} = 0.44$ and ground state momentum $\mathbf{K} = (\frac{3\pi}{4}, 0)$ at fixed interaction strength $V/t = 3.59$. Solid lines represent backflow-inclusive ansatzes, including equivariant and invariant neural networks. Dashed lines show no-backflow ansatzes. The inset provides a zoomed-in view of results without the mean-field Slater ansatz, showing the better performance of the equivariant CNN-based network in achieving lower energies. Error bars represent the corresponding standard deviations.